\documentclass[12pt]{article}
\usepackage{a4wide}
\usepackage{amssymb}
\usepackage{graphicx,color}
\begin{document}
{\renewcommand{\thefootnote}{\fnsymbol{footnote}}
\begin{center}
{\LARGE  Quantum Higgs Inflation}\\
\vspace{1.5em}
Martin Bojowald,$^1$\footnote{e-mail address: {\tt bojowald@gravity.psu.edu}}
Suddhasattwa Brahma,$^2$\footnote{suddhasattwa.brahma@gmail.com}
Sean Crowe,$^3$\footnote{sean.crowe.92@gmail.com}\\
Ding Ding$^1$\footnote{dud79@psu.edu}
and Joseph McCracken$^4$\footnote{jm2264@cornell.edu}
\\
\vspace{0.5em}
$^1$ Institute for Gravitation and the Cosmos,
The Pennsylvania State
University,\\
104 Davey Lab, University Park, PA 16802, USA\\
\vspace{0.5em}
$^2$ Department of Physics, McGill University, Montr\'{e}al, QC H3A 2T8,
Canada\\
\vspace{0.5em}
$^3$ Institute of Theoretical Physics, Jagiellonian University, ul.\ {\L}ojasiewicza 11, 30-348 Krak\'{o}w, Poland\\
and Department of Physics, Georgia Southern University, Savannah, GA 31419 USA\\
\vspace{0.5em}
$^4$ Department of Physics, Cornell University, Ithaca, NY 14853, USA\\
\vspace{1.5em}
\end{center}
}

\setcounter{footnote}{0}

\begin{abstract}
  The Higgs field is an attractive candidate for the inflaton because it is an
  observationally confirmed fundamental scalar field. Importantly, it can be
  modeled by the most general renormalizable scalar potential. However, if the
  classical Higgs potential is used in models of inflation, it is ruled out by
  detailed observations of the cosmic microwave background. Here, a new
  application of non-adiabatic quantum dynamics to cosmological models is
  shown to lead to a multi-field Higgs-like potential, consistent with
  observations of a slightly red-tilted power spectrum and a small
  tensor-to-scalar ratio, without requiring non-standard ingredients. These
  methods naturally lead to novel effects in the beginning of inflation,
  circumventing common fine-tuning issues by an application of uncertainty
  relations to estimate the initial quantum fluctuations in the early
  universe.  Moreover, inflation ends smoothly as a consequence of the derived
  multi-field interactions.
\end{abstract}

\section{Introduction}

One of the most attractive features of cosmic inflation is that it may be able
to explain how the observed large-scale structure of the universe, captured in
the distributions of microwave background radiation or galaxies, could have
evolved out of tiny initial quantum fluctuations
\cite{GuthPi,PowerLargeScales}. It presents a stunning example of how the
microscopic and macroscopic realms can be bridged, potentially allowing
cosmologists to test quantum mechanics through large-scale
observations. However, traditional models of cosmic inflation require the
presence of a certain scalar field in the early universe, the inflaton, with
specific self-interactions that imply negative pressure pervading the entire
early universe. Interactions that reliably imply this behavior often need to
be finely tuned to achieve observationally viable models. Moreover, it is
usually unclear whether ingredients required for fine-tuning can be derived
from fundamental physics.

Here, we present further evidence for the general picture painted by cosmic
inflation by introducing a new combination of cosmological equations with
non-adiabatic methods for quantum dynamics. The novelty of this work is to
capture non-trivial quantum effects in an initial phase of inflation
which have generally been missed in previous studies that considered a single
scalar field on an expanding background. We will show how a quantum state
actively models the self-interactions of a multi-component inflaton field, and
how this new potential can overcome several conceptual and phenomenological
issues encountered when one tries to build inflation on properties of the
Higgs field. The resulting scenario is consistent with current
observations. With more precise future data, it may be used to deduce
properties of the quantum state in the very early universe.

We begin with a Higgs-like field $\psi$ with classical potential
\begin{equation} \label{Vclass}
 V_{\rm class}(\psi)=  M^4\left(1-2\psi^2/v^2+\psi^4/v^4\right)
\end{equation}
where $M$ and $v$ are constants, assumed positive.  When used in a quantum
field theory, this potential is the most general one (up to adding a constant)
that results in a renormalizable theory. It is therefore preferred in a model
of the early universe because it implies physical effects independent of
poorly understood high-energy phenomena. This decoupling allows inflation to
be applied as a low-energy effective theory, avoiding details of quantum
gravity. (Alternatively, one could consider only potentials restricted by a
high-energy theory, for instance in terms of swampland conjectures
\cite{Swamp}.)

For $\psi$ to play the role of a phenomenologically viable inflaton
\cite{PlanckPot}, the potential must be extended in some way, for instance by
including higher-order monomials in $\psi$ or even non-polynomial functions as
in Starobinsky-type inflation \cite{Starobinsky}, or by introducing
non-trivial interactions between $\psi$ and space-time curvature
\cite{HiggsNonMin,HiggsInflation}.

Another possibility is to consider multi-field inflation, in which $\psi$ is
just one of several interacting scalar fields. Additional constants then
appear in potentials and interaction terms, and the system seems to become
more ambiguous as well as more distant from fundamental physics. The new
mechanism presented here, by contrast, will use basic properties of quantum
mechanics to turn any single-field potential, such as (\ref{Vclass}), into an
equivalent multi-field system. All coupling constants between the fields can
then be derived from $v$ and $M$ in (\ref{Vclass}) as well as
fundamental constants such as $\hbar$, but they will also depend on parameters
that characterize the quantum state of $\psi$, such as its fluctuations. The
resulting multi-field inflation is therefore much more restricted than in
usual cosmological model building, in which interaction terms are postulated
so as to obey phenomenological constraints. As one of our main results, the
new quantum-based multi-field scenario is nevertheless consistent with current
observations, without excessive fine-tuning.

\section{Effective potentials for non-adiabatic quantum dynamics}

Heuristically, quantum mechanics always implies that a single classical
variable, such as the position $x$ of a particle or our field $\psi$, is
described by an infinite number of quantum degrees of freedom. For instance,
an initial wave function $\Psi(x)$ can be chosen to be centered at any value
$x_0=\langle\Psi|\hat{x}|\Psi\rangle$, and independently have an arbitrary
variance $(\Delta x)^2=\langle\Psi|(\hat{x}-x_0)^2|\Psi\rangle$ around
$x_0$. Similarly, higher moments $\Delta
x^n=\langle\Psi|(\hat{x}-x_0)^n|\Psi\rangle$ are independent, amounting to
infinitely many quantum degrees of freedom. There are further moments
involving the momentum $p$, or combinations of $x$ and $p$. In standard
quantum mechanics, all these values are captured by a wavefunction or density
matrix.

As time changes, the initial wave function evolves such that, generically, the
moments depend on time. They form an infinite-dimensional dynamical system
with interacting degrees of freedom in addition to the classical $x$. It has
been known for some time that there are equivalent classical-type systems that
describe the same quantum dynamics, derived from multi-component effective
potentials. In particular a semiclassical approximation in which only moments
of second order are considered --- position and momentum variances as well as
their covariance --- has been formulated several times independently
\cite{VariationalEffAc,GaussianDyn,EnvQuantumChaos,QHDTunneling,CQC,CQCFieldsHom}:
With $s=\Delta x$ a single quantum degree of freedom to this order, the
effective potential
\begin{equation}\label{Vs}  
V_{\rm eff}(x,s) = V(x)+ \frac{U}{2s^2}+ \frac{1}{2} V''(x) s^2 
\end{equation}
describes Hamiltonian dynamics equivalent to a first-order approximation in
$\hbar$ of the expectation value $x\approx \langle\hat{x}\rangle$ and
fluctuation $s$ derived from a solution of the Schr\"odinger equation with
potential $V(\hat{x})$. The constant $U$ does not depend on time but only on
the initial state. It obeys $U\geq \hbar^2/4$ as a consequence of Heisenberg's
uncertainty relation. In our application to cosmology, $s$ will be a
multi-field partner to the Higgs-like $\psi$.

The presence of new quantum degrees of freedom in an effective potential may
be unfamiliar to particle physicists and cosmologists, but it is a common
implication of non-adiabatic dynamics. It is possible to derive an effective
low-energy potential from (\ref{Vs}) by minimizing $V_{\rm eff}(x,s)$ with
respect to $s$, at fixed $x$. The result, $V_{\rm low-energy}(x)=V(x)+
\sqrt{UV''(x)}$, is a quantum-corrected potential which contains higher
derivatives of $V$ instead of additional degrees of freedom. Also in quantum
field theory, the low-energy effective potential commonly used in high-energy
physics and cosmology can be seen as a leading-order derivative expansion of a
multi-field potential analogous to (\ref{Vs}) \cite{CW}. (The field theory
version of $\sqrt{UV''(x)}$ quoted here is, for $U=\hbar^2/4$, equivalent to
the Coleman--Weinberg potential \cite{ColemanWeinberg} integrated over the
time component of the wave number. The applicability of these methods to the
relation between quantum field theory and background quantum mechanics has
been demonstrated in \cite{MiniSup}.)  

Because multi-field potentials do not implement a derivative expansion, they
allow studies of non-adiabatic phenomena; see Appendix~\ref{a:NonAd}.
Independent quantum degrees of freedom such as $s$ are relevant whenever the
dynamics is non-adiabatic. Inflation is usually presented in a slow-roll
regime, in which the inflaton changes slowly and its potential energy
dominates over its kinetic energy. The potential then acts as a medium with
tension, implying negative pressure. Such a regime should be well-described by
adiabatic dynamics, and low-energy effective potentials should be sufficient,
as often used in this context. However, as we will show in detail, a
non-adiabatic precursor phase in a multi-field potential such as (\ref{Vs}) is
nevertheless important because it can help to set correct initial
conditions for subsequent long-term inflation without excessive fine-tuning. A
final non-adiabatic phase then helps to end inflation.

The initial stages of inflation are expected to be dominated by quantum
phenomena. A leading-order semiclassical approximation as in (\ref{Vs}) is
then insufficient. The canonical formulation of second-order moments has
recently been extended to higher orders, with complete parameterizations up to
fourth-order moments \cite{Bosonize,EffPotRealize}. Each additional moment
order implies new quantum degrees of freedom, given by three values
$\varphi_1$, $\varphi_2$ and $\varphi_3$ at third order and five at fourth
order. The variance is the sum of the squares of these variables, $\Delta
\psi^2= \varphi_1^2+\varphi_2^2+\varphi_3^2+\cdots$, while third and fourth
order moments of $\psi$ are polynomials of degree three and four,
respectively. The higher-order extension of (\ref{Vs}) is linear in
$\Delta\psi^n$ with coefficients given by the Taylor expansion of the
classical potential around the expectation value, just as seen in (\ref{Vs})
at second order.

In order to obtain a manageable system, we will assume that one of the quantum
variables, called $\varphi$, is most relevant and replaces $s$ in the
mechanics example, such that $\Delta\psi^2= \varphi^2$.  For higher moments,
$\Delta\psi^3=\alpha_3$ and $\Delta\psi^4=\alpha_4\varphi^4$ with parameters
$\alpha_3$ and $\alpha_4$ that describe properties of the quantum state.  Such
an approximation is known as a moment closure, a common technique for coupled
or partial differential equations in fields where there is experimental input
\cite{Closure}. Since our classical potential is quartic, only moments up to
fourth order appear in the effective potential and we do not need higher
orders. For a Gaussian state, $\alpha_3=0$ and $\alpha_4=3$.  The parameters
$\alpha_3$ and $\delta=\alpha_4-3$ may therefore be considered
non-Gaussianities of the background state of $\psi$.

Before we apply the effective potential (\ref{Vs}) to cosmology, we have to
make a small adjustment because the energy contributions of a scalar field
depend on the scale factor $a$ of expanding space. In (\ref{Vs}), as its
derivation shows, the contribution $\frac{1}{2}V''(x)s^2$ results from the
classical potential energy, while $U/(2s^2)$ is a contribution from the
kinetic energy to the effective potential. (The canonical formulation of
moments can be seen to imply $\Delta p^2=p_s^2+U/s^2$ where $p_s$ is the
momentum of $s$. The term $p_s^2$ provides the usual kinetic energy of the new
quantum variable, $s$ or $\varphi$.) The kinetic energy of a free scalar field
in an expanding universe decreases with $a$ because of dilution, while
potential energy, $V(\psi)a^3$, acts like a medium with tension and increases
with $a$. The precise dependence on $a$ can be derived from the canonical
formulation, and leads to an additional factor of $a^{-6}$ in the $U$-term in
an effective potential. It is accompanied by a parameter $V_0$ which, by
definition of a homogeneous model, determines the comoving scale over which
inhomogeneities can be ignored \cite{MiniSup}. The volume $V_0a^3$ should be
larger than the Planck scale in order to avoid the trans-Planckian problem
\cite{TransPlanck,TransPlanck2,TransPlanck3}.  With our ansatz for moments and
defining $\varphi_c^2:=\frac{1}{3}v^2$, the effective potential is
\begin{eqnarray}\label{v-eff}
 V_{\rm eff}
&=& M^4\left(1+2\left(\frac{\varphi^2-\varphi_c^2}{\varphi_c^2}\right) \frac{\psi^2}{v^2}
+\frac{4 \alpha_3\, \psi}{v^4} + \frac{\psi^4}{v^4}
-\frac{2}{3}\frac{\varphi^2}{\varphi_c^2}
+\alpha_{4}\frac{\varphi^4}{v^4}\right) + \frac{U}{2 a^6V_0^2 \varphi^2}\,.
\end{eqnarray}

\section{Cosmological implications}

After a few $e$-folds of inflation, the last term in (\ref{v-eff}) can be
ignored since $a$ grows quickly. It is nevertheless important because it
implies a repulsive potential for $\varphi$, necessitating $\varphi$ to start
out at large values. This alleviates the need to fine-tune the usual initial
condition $\varphi > \varphi_c$ \cite{HybridInfl}. Because $\varphi$ is
initially large, the quartic $\varphi$-term in (\ref{v-eff}) dominates at
early times, along with the $U$-term. Their sum has a local minimum at
$\varphi=\sqrt[^6]{Uv^4/(4a^6V_0^2M^4\alpha_4)}$. Using the Planckian lower
bound for $V_0a^3$, we obtain an upper bound $\varphi_{\rm ini}<
\ell_P^{-1}\sqrt[^6]{Uv^4/(4\alpha_4M^4)}$ on the initial fluctuation
variable. If we assume $v\sim \mathcal{O}(M_{\rm Pl})$ and $M^4\ll M_{\rm Pl}$
as in the detailed analysis to follow, this upper bound is well beyond
$\varphi_c$. (We are using units such that $M_{\rm Pl} = \hbar =c=1$, turning
$M$ into an energy scale.)

\begin{figure}
	\centering
	\includegraphics[width=0.7\textwidth]{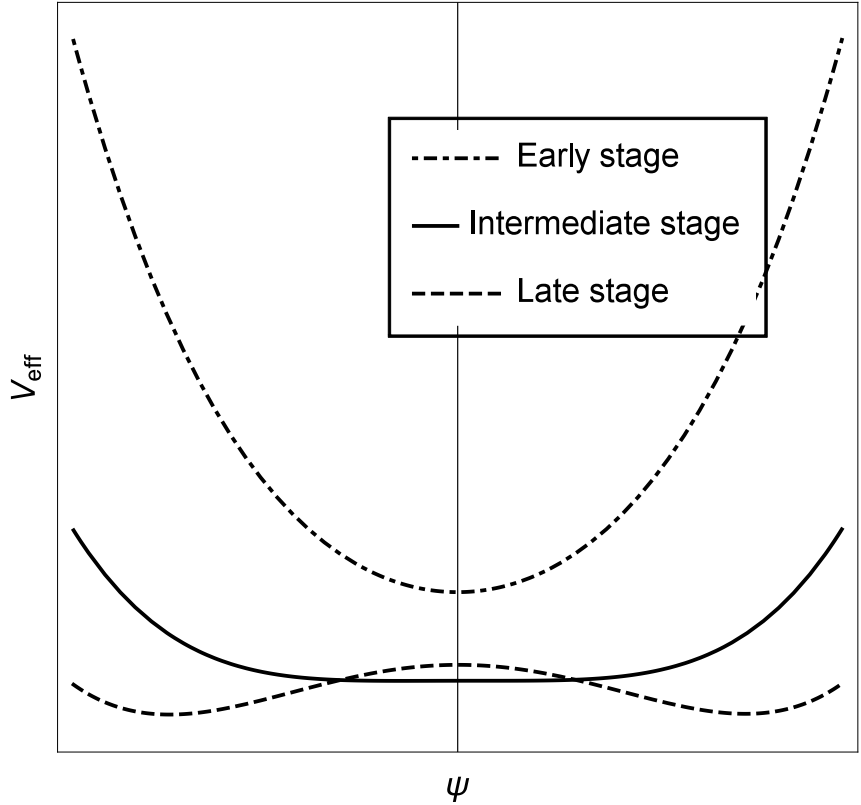}
	\caption{Build-up of a tachyonic potential for the Higgs-like field
          $\psi$. Pre-inflation stage is characterized by $\varphi>\varphi_c$, while
          $\varphi<\varphi_c$ signals the start of the intermediate
          stage.
		\label{phase-psi}}
\end{figure}

\begin{figure}
	\centering
	\includegraphics[width=0.7\textwidth]{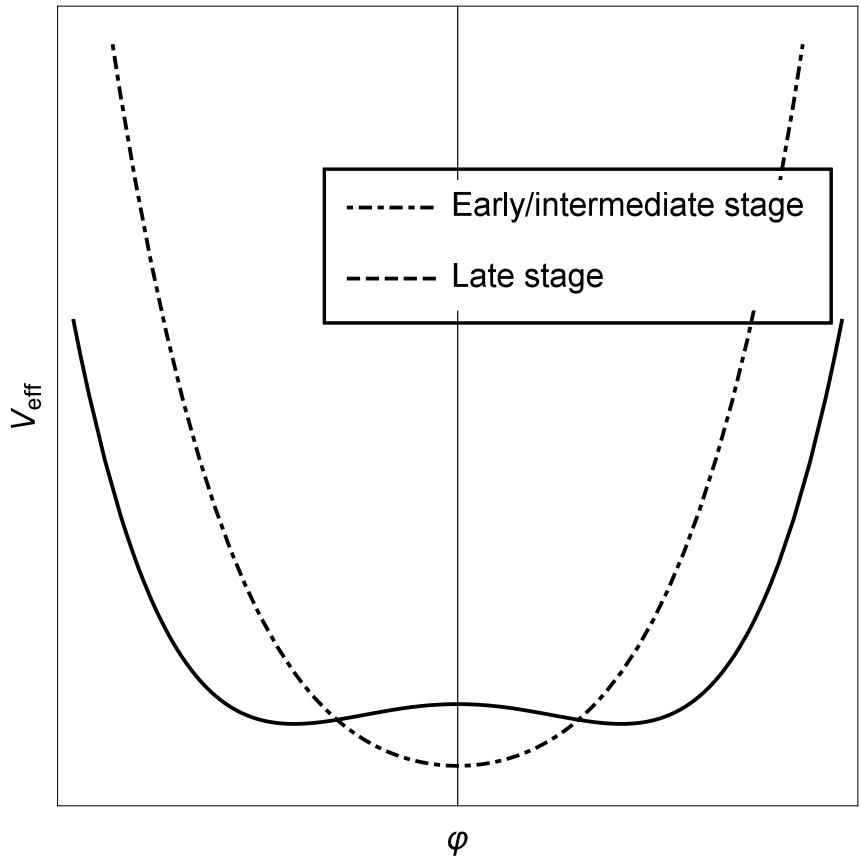}
	\caption{Dynamical restoration of symmetry for the fluctuation field
		$\varphi$. The potential $V_{\rm eff}(\varphi)$ of early stages look similar
  to the intermediate stage.} 
		\label{phase-phi}
\end{figure}

\begin{figure}
	\centering
	\includegraphics[width=0.7\textwidth]{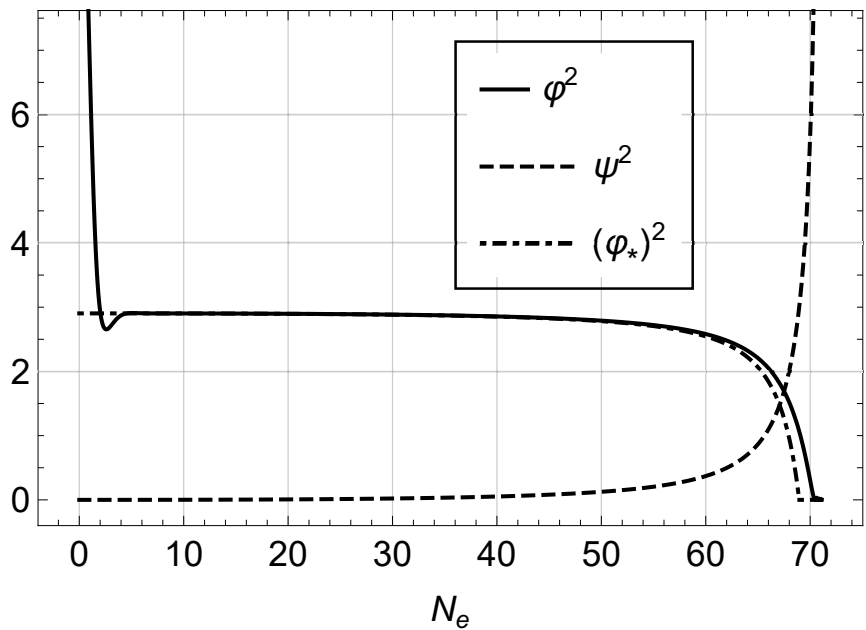}
	\caption{The squared values of exact solutions $\varphi(N_e)$,
          $\psi(N_e)$ 
          and $\varphi_*(N_e)$ are plotted using $v=3$, $\alpha_3=0.05$,
          $\delta=0.1$. The field $\varphi$ follows its minimum closely
          throughout the majority of inflation, while the field $\psi$
	  rolls down to its new potential minimum $\psi_{\rm min}^2=v^2$
	  in the end
	  (see also Fig.~\ref{phase-psi}).
	  Non-adiabatic
          evolution is apparent during the start and end of inflation.
          \label{evo}}
\end{figure}

Our potential (\ref{v-eff}), combining the dynamics of the classical field and
its fluctuation, is of the hybrid-inflation type. These models typically
produce a blue-shifted tilt when one starts with a large $\varphi$ and small
$\psi$, relying on a constant vacuum energy of $\psi$
\cite{HybridInfl}. However, if instead a \textit{waterfall} regime is
responsible for a significant number of $e$-folds, where $\varphi$ stays close
to a local minimum, one may obtain a red tilt for a wide range of parameters
\cite{Waterfall,WaterfallNumerical}. In our case, as opposed to the
traditional hybrid model, the inclusion of non-adiabatic dynamics implies two
phase transitions and the majority of $e$-folds are created in between.  Other
variations of hybrid models \cite{ModHybrid} which produce a red tilt require
additional mechanisms for stability against quantum corrections
\cite{Stability}, making them much less natural by introducing further
tunings. Our model is stable because it describes a quantization of the
generic renormalizable potential (\ref{Vclass}).

As in the original hybrid model, we start with some $\varphi>\varphi_c$, in
accordance with our analysis of the last term in (\ref{v-eff}).  By
construction, $\varphi$ describes quantum fluctuations of $\psi$. It should
indeed be large for a highly quantum initial state, even while the expectation
value $\psi$ remains small at the local minimum of its early-time potential
Fig.~\ref{phase-psi}. For such a large value of $\varphi$, its early-time
potential, shown in Fig.~\ref{phase-phi}, is steep. The field quickly
approaches one of its minima driven by the $\varphi^4$-term in (\ref{v-eff}).

Once $\varphi$ crosses $\varphi_c$, the potential of $\psi$,
Fig.~\ref{phase-psi}, changes to its tachyonic intermediate-stage form with
true minima located at non-zero $\psi$.  In (\ref{v-eff}), reflection symmetry
of $\psi$ is broken for any non-zero $\alpha_3$. As the tachyonic contribution
builds up, the field starts slowly rolling away from the origin, acting as the
instability required to kick-start the waterfall regime. This slow change, now
in an adiabatic phase, enables $\varphi$ to follow its vacuum expectation
value, $\varphi_*$. Eventually, $\varphi_*\rightarrow 0$, as indicated by the
late-time potential, Fig.~\ref{phase-phi}. This second phase transition
restores reflection symmetry for $\varphi$. In summary, $\varphi$ causes the
traditional phase transition when it crosses $\varphi_c$, and the subsequent
slow roll of $\psi$ down its tachyonic hilltop eventually triggers a second
phase transition. The adiabatic phase of slow-roll inflation takes place
between the phase transitions.

For quantitative predictions, we solve the full equations
\begin{eqnarray}
	\label{eom-psi}
	\ddot{\psi}+3H\dot{\psi}&=& 
	 M^4\left(-\frac{4\psi}{v^2}
	\left(\frac{\varphi^2-\varphi_c^2}{\varphi_c^2}+\frac{\psi^2}{v^2}\right)
	-\frac{4\alpha_3}{v^4}\right)\\
	\ddot{\varphi}+3H\dot{\varphi}&=&  
	M^4\left(\frac{4\varphi}{3\varphi_c^2}
	\left(1-\frac{3\psi^2}{v^2}\right)-\frac{4\alpha_4\varphi^3}{v^4}\right)\\ 
        6H^2&=&  \dot{\psi}^2+\dot{\varphi}^2
	+2V_{\rm eff}(\psi,\varphi) 
\end{eqnarray}
using numerics. (A dot represents a derivative by proper time.) In
Fig.~\ref{evo}, we show $\varphi$ and $\psi$ as functions of the number of
$e$-folds, $N_e\equiv \ln (a(t)/a(0))$. Non-adiabatic dynamics is visible in
both the beginning and end stages of inflation, caused by the large departure
of the fluctuation field $\varphi$ from its minima, $\varphi_*\equiv\pm
v\sqrt{\alpha_4^{-1}(1-3\psi^2/v^2)}$ (early stage) or $\varphi_*=0$ (late
stage). Note that $\varphi_*^2\approx
\varphi_c^2+\mathcal{O}(\psi^2/v^2,\delta)$. This value is large (Planckian),
but it determines the local minimum of the potential where its value, of the
order $M^4$, is sub-Planckian. The dynamics is therefore not affected by
quantum gravity.

\begin{figure}
	\centering
	\includegraphics[width=0.8\textwidth]{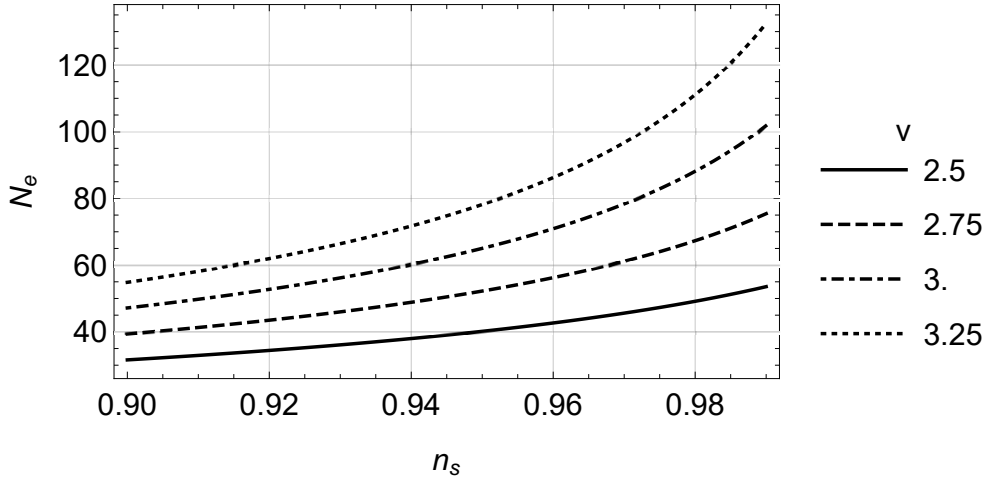}
	\caption{Number of $e$-folds, $N_e$,  as a function of the
          spectral index $n_{\rm s}$, using the approximation
		(\ref{N-ns}) with various $v$. 
		\label{N-ns-fig}}
\end{figure}

It is possible to derive analytical expressions for observables using a
slow-roll approximation combined with small non-Gaussianity, $\delta$ and
$\alpha_3$. Since the initial $\psi$ is small near its minimum, we may ignore
the $\psi^3$-term in (\ref{eom-psi}) to obtain the spectral index $n_{\rm s}$
at Hubble exit. Since inflation ends shortly after $\psi^2$ grows to
$\psi^2=v^2/3$, we have $\varphi_*=0$. Assuming $\varphi^2\approx\varphi_*^2$
and small non-Gaussianity, we
derive, as detailed in Appendix~\ref{a:Cosmo},
\begin{equation}\label{N-ns}
	n_{\rm s}\approx 1-12\frac{\delta}{\alpha_4v^2}\quad,\quad
	N_e\approx
        \frac{f(1-n_{\rm s},v,\alpha_3)}{1-n_{\rm s}}
\end{equation}
with a specific but lengthy function $f$.  In non-minimal Higgs inflation,
$f(1-n_{\rm s},v,\alpha_3)\approx 2$ is constant \cite{HiggsNonMin} while here
it increases logarithmically with growing $1-n_{\rm s}$ and typically ranges
from $1\lesssim f(1-n_{\rm s},v,\alpha_3)\lesssim 5$.  The second equation
in (\ref{N-ns}) is plotted in Fig.~\ref{N-ns-fig} for $\alpha_3=0.05$.

\begin{figure}
	\centering
	\includegraphics[width=0.8\textwidth]{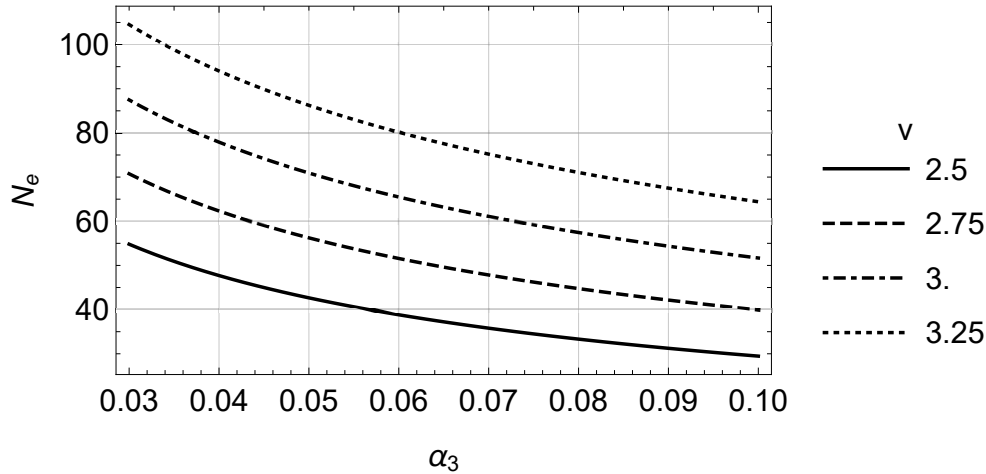}
	\caption{Number of $e$-folds, $N_e$, as a function of the
          non-Gaussianity parameter $\alpha_3$, using (\ref{N-ns}) and
          assuming $n_{\rm s}\approx 0.96$. 
Non-Gaussianity speeds up the departure
		from adiabatic evolution, ending inflation
                earlier. Here, $\delta=0.1$. For smaller
                non-Gaussianity parameters $\delta$, $N_e$ increases because
                $n_{\rm s}$, a function of the ratio
                $\delta/\alpha_4\approx \delta/3$ is then closer to one. 
		\label{N-alpha3}}
\end{figure}

\section{Conclusions}

Aside from the parameter $v$ that appears in common Higgs-like or hybrid
models, our observables depend on two new parameters, $\alpha_3$ and $\delta$,
related to the quantum state. They describe non-Gaussianity of the background
field (as opposed to perturbations) and effectively control the amount of
non-adiabatic evolution due to its modulation of the shifted local
$\varphi$-minima at $\varphi_*$. The dependence of the total number of
$e$-folds on $\alpha_3$ is shown in Fig.~\ref{N-alpha3}, using the analytical
solutions. Observational requirements are consistent with a nearly Gaussian
state. The different parameter values reveal that the hierarchy in our set of
parameters is much more rigid than in traditional hybrid model, leaving less
room for tuning and ambiguity and making our results more robust. In the
traditional case there are three independent parameters, while we have only
two, inherited from a single-field model accentuated by its quantum
fluctuations. Importantly, having a true single-field model masquerading as a
two-field one allows us to avoid fine-tuning issues known for small-field
hilltop models and yet have a small tensor-to-scalar ratio $r$ well-within
observable bounds. As revealed by numerics, the small $r$ is implied by a
small slow-roll parameter $\epsilon$ of the adiabatic field (a combination of
both $\psi$ and $\varphi$ responsible for the curvature perturbation).

\section*{Acknowledgements}
SB thanks Robert Brandenberger for discussions at an early stage of this project.
This work was supported in part by NSF grant PHY-1912168. SB is supported in
part by a McGill Space Institute Fellowship and a CITA National Fellowship. SC is
supported by the Sonata Bis Grant No. DEC-2017/26/E/ST2/00763 of the National
Science Centre Poland.

\begin{appendix}
\section{Non-adiabatic methods} 
\label{a:NonAd}

Non-adiabatic methods, used in our letter in order to derive effective
potentials with new quantum degrees of freedom, can be interpreted as an
extension of Gaussian methods to a more general class of possibly non-Gaussian
states. In addition, they can be formulated as a systematic semiclassical
expansion. As a starting point, we consider the time-dependent variational
principle with quantum action
\begin{equation} \label{S}
S=\int {\rm d}t \left \langle \Psi
  \left|\left(i \partial_t-\hat{H}\right)\right| \Psi \right
\rangle\,,
\end{equation}
specialized to a Gaussian state
\begin{equation} 
\Psi_{\psi, p_{\psi},\varphi, p_{\varphi}}(\bar{\psi})=\frac{1}{\left(2
    \pi \varphi^2\right)^{1/4}}
\exp\left(-{\textstyle\frac{1}{4}}\varphi^{-2}(1- 2
i \varphi  p_{\varphi})(\bar{\psi}-\psi)^2\right)
\exp(i p_{\psi}
  (\bar{\psi}-\psi))\exp(-{\textstyle\frac{1}{2}}i \varphi p_{\varphi}
  ) \label{Psi}
\end{equation}
with parameters $\psi$, $p_{\psi}$, $\varphi$ and $p_{\varphi}$. Using the
chain rule for the time derivative in (\ref{S}) and explicit expectation
values in $\Psi$, the action can be seen to equal 
\begin{equation}
  S=\int {\rm d}t \left(\dot{\psi}p_{\psi}+\dot{\varphi}p_{\varphi}-H_{\rm
      G}\right) 
\end{equation}
with the Gaussian effective Hamiltonian $H_{\rm
  G}(\psi, p_{\psi},\varphi,
p_{\varphi})=\langle\Psi|\hat{H}|\Psi\rangle_{\psi, p_{\psi},\varphi,
  p_{\varphi}}$. This class of states is therefore described by two
independent canonical degrees of freedom, $\psi$ which equals the expectation
value of $\hat{\psi}$ as well as $\varphi$ which equals the quantum fluctuation
$\Delta \psi$ of a Gaussian state.

These properties are not restricted to Gaussian states but always hold at the
semiclassical level to first order in $\hbar$. First, the expectation value of
the Hamilton operator $\hat{H}$ always provides a quantum Hamiltonian of an
equivalent dynamical system. One can show this systematically using the
Poisson methods of \cite{EffAc,Karpacz}, which first define a quantum Poisson
bracket
\begin{equation}\label{Poisson}
 \{\langle\hat{A}\rangle,\langle\hat{B}\rangle\}=
 \frac{\langle[\hat{A},\hat{B}]\rangle}{i\hbar}
\end{equation}
for expectation values of any pair of operators, $\hat{A}$ and
$\hat{B}$. Using the Leibniz or product rule, this Poisson bracket is defined
for all moments of a state, including basic expectation values
$\langle\hat{\psi}\rangle$, quantum fluctuations $\Delta\psi$ and $\Delta
p_{\psi}$, as well as higher moments. If $\hat{B}=\hat{H}$ is taken to be the
Hamilton operator, (\ref{Poisson}) shows that Hamilton's equations of
$\langle\hat{H}\rangle$ for the moments are equivalent to Heisenberg's
equation for operators.

Secondly, (\ref{Poisson}) applied to moments shows that they imply independent
phase-space variables. However, moments do not directly appear in canonical
form for this bracket, as shown for instance by $\{(\Delta\psi)^2,(\Delta
p_{\psi})^2\}= 4 C_{\psi p_{\psi}}$ where $C_{\psi p_{\psi}}$ is the
covariance. Nevertheless, as shown independently in various contexts
\cite{VariationalEffAc,GaussianDyn,EnvQuantumChaos,QHDTunneling,CQC,CQCFieldsHom},
there is an invertible transformation,
\begin{eqnarray}
 (\Delta \psi)^2 &=& s^2 \label{s}\\
 C_{\psi p_{\psi}} &=& s p_s\\
 (\Delta p_{\psi})^2 &=& p_s^2+ \frac{U}{s^2}\,, \label{p}
\end{eqnarray}
that maps second-order moments to a canonical pair, $(s,p_s)$, together with a
conserved quantity $U$. Heisenberg's uncertainty relation implies $U\geq
\hbar^2/4$. The Gaussian restriction amounts to $U=\hbar^2/4$ to this order.

The effective Hamiltonian $\langle\hat{H}\rangle$ can the be written entirely
in terms of canonical variables. The kinetic energy, $\frac{1}{2}
\hat{p}_{\psi}^2$, contributes three terms, using
$\langle\hat{p}_{\psi}^2\rangle=\langle\hat{p}_{\psi}\rangle^2+ (\Delta
p_{\psi})^2$ and (\ref{p}): We now have two contributions to the effective
kinetic energy, $\frac{1}{2}\langle\hat{p}_{\psi}\rangle^2+ \frac{1}{2}p_s^2$,
and a momentum-independent contribution $\frac{1}{2}U/s^2$ which we can
combine with $\langle V(\hat{\psi})\rangle$ in an effective potential. The
latter contribution can be written in terms of second-order moments after
using a Taylor expansion of $\langle V(\hat{\psi})\rangle= \langle
V(\langle\hat{\psi}\rangle+ \Delta\hat{\psi})$ in
$\Delta\hat{\psi}=\hat{\psi}-\langle\hat{\psi}\rangle$:
\begin{equation}
\langle V(\hat{\psi})\rangle=
V(\langle\hat{\psi}\rangle) +\frac{1}{2} V''(\langle\hat{\psi}\rangle)
(\Delta\psi)^2+\cdots
\end{equation} 
(The Taylor expansion in an operator, $\Delta\hat{\psi}$, is formal for a
general potential, but merely presents an efficient way of writing $\langle
V(\hat{\psi})\rangle$ in terms of central moments if the potential is
polynomial, as in our case.) Using (\ref{s}), the combined effective potential
is
\begin{equation}\label{V}
 V_{\rm eff}(\psi,s)= \frac{U}{2s^2}+ V(\psi)+ \frac{1}{2}V''(\psi)s^2+\cdots
\end{equation}
where we suppressed expectation-value brackets around $\psi$.

As shown in \cite{Bosonize,EffPotRealize} and described in our letter,
canonical variables for moments can also be found at higher orders, based on
the general Casimir--Darboux theorem of Poisson geometry. The resulting
quantum dynamics is non-adiabatic because moments evolve as described by
independent degrees of freedom. The relationship to adiabatic methods can be
seen by applying an adiabatic approximation within our setting in which
moments, such as $s$, are assumed to evolve slowly compared with the basic
expectation values, $\psi$ and $p_{\psi}$. To leading order in an adiabatic
approximation, the usual low-energy effective potential is obtained
\cite{EffAc}. Here, it is sufficient to minimize (\ref{V}) with respect to $s$
(at fixed $\psi$), such that $s_{\rm min}=\sqrt[4]{U/V''(\psi)}$ and $V_{\rm
  eff}(\psi,s_{\rm min})= V(\psi)+ \frac{1}{2}\hbar/\sqrt{V''(\psi)}$
(assuming $U=\hbar^2/4$). This potential can be seen to equal the
quantum-mechanics version of the low-energy or Coleman--Weinberg potential
\cite{ColemanWeinberg}. Also the field-theory version of this potential can be
rederived from moments \cite{CW}, where it is again recognized as the
adiabatic approximation of a more general non-adiabatic description. The
methods used here are therefore applicable very broadly, including in
situations of quantum field theory relevant for the early universe.

\section{Derivation of cosmological parameters} 
\label{a:Cosmo}

The cosmological parameters presented in our letter follow, as in
\cite{Waterfall}, from a slow-roll approximation that implements the
conditions $\psi^2\ll v^2$ as long as $\varphi^2\approx \varphi^2_*$ is near a
local minimum, identified as the dominant slow-roll regime in our
model. Moreover, we have $\ddot{\varphi}\ll 3H\dot{\varphi}$ and
$\dot{\varphi}^2\ll V$. Under slow-roll, the equations of motion then read
\begin{eqnarray} \label{phi-eom}
 \frac{3H\dot{\varphi}}{M^4}&=&
 \frac{4\varphi}{3\varphi_c^2}\left(1-\frac{3\psi^2}{v^2}\right) 
 -\frac{4\alpha_4\varphi^3}{v^4}\\
 \label{psi-eom}
 \frac{3H\dot{\psi}}{M^4}&=& -\frac{4\psi}{v^2}
 \left(\frac{\varphi^2-\varphi_c^2}{\varphi_c^2}+ \frac{\psi^2}{v^2}\right)- 
\frac{4\alpha_3}{v^4}\,. 
\end{eqnarray}
We present here a simplified derivation that shows the key features. Further
details can be found in \cite{EffPotInflation}.

Using the equations above, one can calculate
$\epsilon_{\sigma}=\epsilon_{\psi}+\epsilon_{\varphi}$ the standard two-field
slow-roll parameter for the effective adiabatic field \cite{Waterfall}.  Using
$\alpha_4=3+\delta$, $\varphi\approx\varphi_*\approx \varphi_c(1-\delta/6)$
and $\psi\approx 0$ at early times of Hubble exit, we find
\begin{eqnarray}
\epsilon_{\varphi} &\equiv& \frac{1}{2}\left(\frac{V_{\varphi}}{V}\right)^2
\approx 0+O(\delta^2)\\
\epsilon_{\psi} &\equiv& \frac{1}{2}\left(\frac{V_{\psi}}{V}\right)^2
\approx 0+O(\alpha_3^2)\,.
\end{eqnarray}
With the value used in the letter, $\alpha_3\sim\delta\sim 0.05$, we
see that the constraint
\begin{equation}
	r = 16 \epsilon_{\sigma} < 0.07
\end{equation}
tensor-to-scalar ratio can easily be satisfied. This has also been checked
numerically \cite{EffPotInflation}.  Setting the power spectrum of curvature
perturbations to its observed value then determines the energy scale of
inflation in our model, $V_{\rm in}$. Given the bound on $\epsilon_{\sigma}$,
the condition 
\begin{eqnarray}
	\mathcal{P}_{\zeta} = \frac{V_{\rm in}}{24 \pi^2
          \epsilon_\sigma} \sim 10^{-9}\,, 
\end{eqnarray}
$V_{\rm in}$ is well below the Planck scale.

The spectral index is given by the general expression
\begin{equation}
 n_{\rm s}= 1-6\epsilon_{\sigma}+2\eta_{\sigma\sigma}
\end{equation}
in terms of the second slow-roll parameter \cite{Waterfall}, in addition to
the first one mentioned above,
\begin{equation} \label{eta}
	\eta_{\sigma\sigma}= \eta_{\varphi\varphi}\cos^2\theta
	+\eta_{\psi\psi}\sin^2\theta
	+2\eta_{\varphi\psi}\sin\theta\cos\theta
\end{equation}
where $\theta$ is such that
\begin{equation}
 \cos\theta= \frac{\dot{\varphi}}{\sqrt{\dot{\varphi}^2+\dot{\psi}^2}}
	\quad,\quad
	\sin\theta= \frac{\dot{\psi}}{\sqrt{\dot{\varphi}^2+\dot{\psi}^2}}\,.
\end{equation}
During waterfall slow-roll, $\varphi$ adiabatically tracks its local,
$\psi$-dependent  minimum at
$\varphi_*^2= v^4(1-3\psi^2/v^2)/(3\alpha_4\varphi_c^2)$. Therefore,
$\dot{\varphi}\approx -3\psi(\alpha_4\varphi_*)^{-1}\dot{\psi}\ll\dot{\psi}$
due to small $\psi$, and
we obtain
\begin{equation}
	\cos\theta
	\approx -\frac{3\psi}{\alpha_4\varphi_*}\sin\theta\quad,\quad
	\sin\theta
	\approx 1\,.
\end{equation}
Moreover, $|\epsilon_\sigma| \ll |\eta_{\sigma\sigma}|$, such that the
dominant contribution to the spectral index is then given by
\begin{equation}
 \eta_{\psi\psi}\sin^2\theta \approx \eta_{\psi\psi}=
 \frac{1}{V}\frac{\partial^2V}{\partial\psi^2}\approx 
-\frac{4\delta M^4}{\alpha_4V_{\rm in}v^2}
\end{equation}
with the initial potential $V_{\rm in}=V(0,\varphi_{\rm c})\approx
M^4(2/3+\delta/9)$.  Therefore,
\begin{equation}
	n_{\rm s}\approx 1-8\frac{\delta M^4}{\alpha_4V_{\rm in}v^2} \approx
1-12\frac{\delta}{\alpha_4v^2}\,. \label{ns}
\end{equation}

Since $\eta_{\psi\psi}$ is the dominant slow-roll parameter in (\ref{eta}), it
determines the end of inflation when it reaches a value of order one. Using
slow-roll solutions, this is the case when $\psi^2=v^2/3$. If we can find
$\psi$ as a function of the number of $e$-folds, $N$, inverting this equation
determines the full number of $e$-folds. During waterfall slow-roll, $\varphi$
is nearly constant at $\varphi^2\approx
(3/\alpha_4)\varphi_c^2-3\psi^2/\alpha_4$ as a consequence of (\ref{phi-eom})
with $\dot{\varphi}\approx 0$. Therefore,
$(\varphi^2-\varphi_c^2)/\varphi_c^2\approx 
-\delta/\alpha_4-3\psi^2/v^2$. 
Using ${\rm d}N=H{\rm d}t$, the
second equation of motion, (\ref{psi-eom}), can then be written as
\begin{equation}
 \frac{{\rm d}\psi}{{\rm d}N}\approx  \frac{M^4}{V_{\rm
     in}}\frac{4\psi}{v^2} 
 \left(\frac{\delta}{\alpha_4}+\frac{2\psi^2}{v^2}\right)
 \approx  \frac{1-n_{\rm s}}{2} \psi+ \frac{8M^4}{V_{\rm in}v^4}\psi^3
\end{equation}
in terms of the number of $e$-folds, $N$. (For now, we ignore the term
$-4\alpha_3/v^4$ in (\ref{psi-eom}).) Rewriting this equation as
\begin{equation}
 (1-n_{\rm s}){\rm d}N\approx
 2\left(\frac{1}{\psi}-\frac{\beta\psi}{1+\beta\psi^2}\right) {\rm d}\psi
\end{equation}
with $\beta= 16M^4/(V_{\rm in}v^4(1-n_{\rm s}))$, we see that at the end
of inflation, marked by $\psi=-v/\sqrt{3}$ (assuming $\alpha_3>0$),
$N$ equals $(1-n_{\rm s})^{-1}$
times a function that depends logarithmically on $1-n_{\rm s}$ and $v$. If
the term $-4\alpha_3/v^4$ in (\ref{psi-eom}) is not ignored, ${\rm d}\psi/{\rm
  d}N$ is still given by a third-degree polynomial in $\psi$ with a more
lengthy factorization, but our general argument about the dependence on
$1-n_{\rm s}$ still holds true, then also implying a logarithmic dependence
on $\alpha_3$.
\end{appendix}


\end{document}